# Oil-US Stock Market Nexus: Some insights about the New Coronavirus Crisis

Claudiu Tiberiu ALBULESCU[a,b], Michel MINA[b], Cornel OROS[b,c*]

[a] *Management Department, Politehnica University of Timisoara, 2, P-ta. Victoriei, 300006, Timisoara, Romania.*

[b] *CRIEF, University of Poitiers, 2, Rue Jean Carbonnier, Bât. A1 (BP 623), 86022, Poitiers, France.*

[c] *LEO, University of Orléans, rue de Blois - BP 26739, 45 067 Orléans, France.*

**Abstract**
We provide a new investigation of the relationship between oil and stock prices in the context of the outbreak of the new coronavirus crisis. Specifically, we assess to what extent the uncertainty induced by COVID-19 affects the interaction between oil and the United States (US) stock markets. To this end, we use a wavelet approach and daily data from February 18, 2020 to August 15, 2020. We identify the lead-lag relationship between oil and stock prices, and the intensity of this relationship at different frequency cycles and moments in time. Our unique findings show that co-movements between oil and stock prices manifest at 3-5-day cycle and are stronger in the first part of March and the second part of April 2020, when oil prices are leading stock prices. The partial wavelet coherence analysis, controlling for the effect of COVID-19 and US economic policy-induced uncertainty, reveals that the coronavirus crisis amplifies the shock propagation between oil and stock prices.

**Keywords**: oil prices, stock prices, new coronavirus, COVID-19, economic policy uncertainty, wavelets, time-frequency domain, partial wavelet analysis

**JEL codes**: Q41, G11, G15, C32.

*Corresponding author. E-mail: cornel.oros@univ-poitiers.fr.

Acknowledgements: *This work was supported by a Grant of the Romanian National Authority for Scientific Research and Innovation, CNCS–UEFISCDI, Project Number PN-III-P1-1.1-TE-2019-0436.*

# 1. Introduction

The new coronavirus (COVID-19) crisis has forced oil and financial markets to run into headwinds. Firstly, the oil price plunged with more than 20% in one single day, on March 9, 2020. Several factors explain the oil price dynamics in the context of COVID-19 crisis. On the one hand, the pandemic crisis generated a worldwide contraction of the real activity, primarily affecting the transport sector, and therefore, the demand for oil. In addition, given the amplitude of its social impact (millions of infected people, high fatality ratios, and the associated blockages of the health care systems), COVID-19 created a lot of uncertainty into the markets. On the other hand, the severe slowdown in the Chinese oil demand recorded in February 2020 and the fall of initial negotiations between the Organization of the Petroleum Exporting Countries (OPEC) and Russia regarding oil production determined international producers to launch a price war on the market. Consequently, the crude oil prices approached to their lowest level during the last decades.

Secondly, the US stock markets recorded price bubbles before the globally propagation of COVID-19. Likewise, over the last five years, the S&P 500 index has reached a maximum of 3,380 points on February 14, 2020. In the context of a booming uncertainty and financial volatility, S&P 500 plummeted 4.4% on February 28. A highly more severe shock was experienced on March 9, 2020, when the oil price crash has transmitted to the stock price dynamics (the so-called "Black Monday").

Given this context, the purpose of our paper is to provide new evidence on the relationship between oil and stock prices, by taking explicitly into consideration the uncertainty induced by the new coronavirus crisis. We investigate thus the intensity of this relationship at different frequency cycles and moments in time.

The mechanisms explaining the relationship between oil price fluctuations and stock market valuations are well established in the literature. On the one hand, the theoretical arguments state that an increase in oil prices leads to a reduction of stock prices (Hamilton, 2009). According to this theory, oil prices act as an inflation tax on consumers and producers, with undesirable effects on anticipated profits and dividends, generating a decreasing trend on stock markets. On the other hand, according to the theory of commodity markets financialization, oil and stock prices co-move (Zhang et al., 2017), whereas the recent positive stocks-oil correlation is explained by both a contraction of global demand and volatility shocks (Bernanke, 2016).

Starting from these theoretical arguments, the empirical literature dealing with oil – stock market nexus reveals mixed findings. Indeed, all types of interactions are documented in the literature: a negative impact of oil shocks on stock markets (e.g. Kilian and Park, 2009; Basher et., 2018), a positive impact (Demirer et al., 2015), or no significant impact (Apergis and Miller, 2009; Bastianin et al., 2016). The only consensual fact provided by this abundant empirical literature suggests that the oil-stock nexus is nonlinear (Reboredo, 2010), time-varying (Ji et al., 2020), asymmetric (Balcilar et al., 2019), and highly sensitive to specific events like for instance the financial crisis (Albulescu et al., 2019; Bouri, 2015; Reboredo and Ugolini, 2016; Roubaud and Arouri, 2018).

The mixed findings reported by the empirical literature can therefore be explained by the manifestation of different forms of uncertainty which significantly influence the relationship between oil prices and stock market returns. First, the economic policy uncertainty negatively affects stock market returns, while it contributes to higher stock market volatility (e.g. Arouri et al., 2016). Second, oil prices are sensitive to geopolitical factors, commercial settlements, and different policy measures (Albulescu et al., 2020). Therefore, extending the literature to the oil context, Aloui et al. (2016) find that higher policy uncertainty is positively connected with oil prices during periods preceding crisis episodes.

Nevertheless, these studies have not explicitly analyzed the interactions between oil and stock markets in presence of specific uncertainty shocks. Integrated analyses dealing with the impact of uncertainty on both the oil and stocks markets are very rare in the literature (few exceptions are represented by the papers of Kang and Ratti (2013), Kang et al. (2017), Fang et al. (2018), Badshah et al. (2019), Albulescu (2021)). Our paper aims to fill the gap in the literature by exploring the co-movements and the nature of the relationships between oil price dynamics and stock markets returns in the presence of the COVID-19 global shock. Indeed, the recent paper by Prabheesh et al. (2020) explores the time-varying dependence between stock and oil prices in the context of the new coronavirus crisis. Different from this work, we resort to a time-frequency framework in order to isolate the effect of COVID-19 on oil-stock price co-movements and, at the same time, to see if co-movements manifest at specific frequency cycles.

Therefore, the second element of originality of our paper is represented by the proposed methodological approach. We apply a wavelet model, combining the time and the frequency domain analyses, to assess simultaneously how variables are related at different frequencies and moments in time. More precisely, we resort to a Continuous Wavelet Transform (CWT) and we first use the wavelet coherence (WTC) to identify the co-movements and the phase-differences (lead-lag situations) between oil and stock prices. As novelty, we control for the impact of coronavirus-induced uncertainty by using the partial wavelet coherence (PWC) proposed by Grinsted et al. (2004) and extended by Mihanović et al. (2009). We consider the coronavirus-induced uncertainty (e.g. the new infection cases at global level and in the US, the global fatality ratio), as well as the policy-induced economic uncertainty (US EPU), for robustness purpose.

## 2. Methodology and data

To filter the data through the wavelet transform, we rely on the CWT. Likewise, for a time series $X_t$ (with $t = 1, \ldots N$) and a discrete sequence $x_t$, the CWT is defined as (Grinsted et al., 2004):

$$W_\tau^X(s) = \sqrt{\frac{\delta t}{s}} \sum_{t=1}^{N} x_t \psi_0 \left[(t-\tau)\frac{\delta t}{s}\right], \tag{1}$$

where $\tau$ is the time scale and $s$ is the wavelet scale.

The co-movements and the lead-lag relationship between two series $X_t$ and $Y_t$ can be analyzed resorting to the WTC approach proposed by Torrence and Webster (1999):

$$R_t^2(\tau, s) = \frac{|S(s^{-1}W_\tau^{XY}(s))|^2}{S|(s^{-1}|W_\tau^X(s)|^2)| \times S|(s^{-1}|W_\tau^Y(s)|^2)|}, \tag{2}$$

where $S \cdot$ is the smoothing parameter that ensure a balance between resolution and significance.

Starting from the WTC, Grinsted et al. (2004) propose the PWTC to remove the effect of a third variable $X'_t$ on the connection between $X_t$ and $Y_t$. Further, building upon Grinsted et al. (2004), Mihanović et al. (2009) advance the PWTC squared, which ranges from 0 to 1, as follows:

$$RP^2(Y, X : X') = \frac{|R(Y,X) - R(Y,X') \cdot R(Y,X)^*|^2}{[1-R(Y,X')]^2 \times [1-R(X,X')]^2}. \tag{3}$$

We use daily data from February 18, 2020 to August 14, 2020, covering both the pre-pandemic and the pandemic phase of COVID-19. The starting date is February 18, 2020, immediately after the peak of the new infection cases reported by China. Daily data are available on the WHO website until August 15, 2020. The oil price data are extracted from the US Energy Information Administration and refers to the West Texas Intermediate (WTI). The S&P 500 data are obtained from Datastream, whereas the US EPU daily data are extracted from Baker et al. (2016) and are daily updated. Finally, the COVID-19 statistics come from World

Health Organization (WHO) situation reports and refer to the total infection cases at global level and in the US, as well as to the fatality ratio (the ratio between reported deaths and infection cases) recorded at global level. WHO's report released at date "t" indicates the COVID-19 numbers reported by countries at "t-1". Therefore, we use one lag for the COVID-19 statistics.

It is well known that geopolitical risks strongly influence oil prices. Starting with February 2020, several round of negotiations between OPEC and Russia take place, in order to establish limit for oil extraction and to prevent huge price drops. However, this ongoing "conflict" certainly affected the nexus between oil and stock prices. Likewise, to remove the effect of geopolitical risks on oil prices we have first orthogonalized the oil prices on a Google trend index showing the global search of the "OPEC" during the analyzed period. We use this proxy for the geopolitical risk because the well-known index proposed by Caldara and Iacoviello (2018) is unavailable since March 11, 2020. In the second step, all variables are expressed in logarithmic difference.

## 3. Results

We start the analysis with the presentation of WTC results. For each figure, the black contour inside the cone of influence (COI) designates the 5% significance level. Y-axis measures frequencies or scales (period on the graph), ranging from the shortest scale (1-4-day cycle) to the longest scale (16-32-day cycle). X-axis represents the time period (126 observations). The blue color inside the COI is associated with low co-movements, while the yellow color, with high co-movements. The area below the COI shows no statistical influence.

Figure 1 shows that oil prices and stock prices co-move in general, supporting thus the financialization theory. Significant co-movements are recorded at short and medium scales, i.e. 2-6-day cycle, and manifest between February 20 and March 10, 2020 and for the period running from June to July 2020.

Fig. 1. Wavelet coherence between WTI and S&P 500

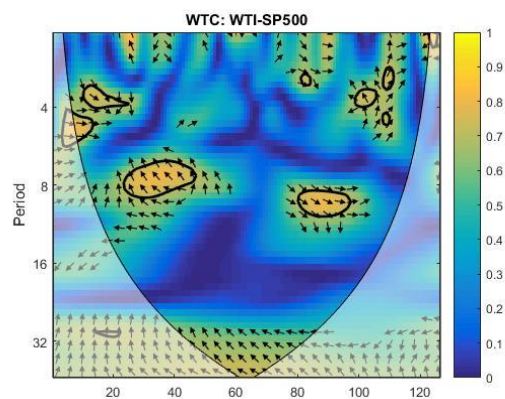

Notes: (i) On the X-axis, observation 1 corresponds to February 18, 2020, observation 60 corresponds to May 12, and observation 120 to August 6; (ii) The arrows indicate the phase differences between the two series. Arrows pointing to the right mean that the variables are in-phase (they co-move), to the right and up mean that the oil prices are leading stock prices, and to the right and down mean that oil prices are lagging. Arrows pointing to the left show that the variables are out-of-phase, to the left and up mean that that oil prices are lagging, and to the left and down mean that oil prices are leading.

It is not clear, however, if the oil prices lead the US stock prices, or the opposite applies. The 2-6-day cycle indicates the average number of days which are necessary for the stock prices to accommodate shocks in oil prices. At the same time, in the medium run (6-8-day cycle) and for the period running from March 20 to April 20, 2020, we clearly notice that the variables are out-of-phase and oil prices are lagging stock prices.

In the second step, we isolate the effects of two types of uncertainties on the oil-US stock price nexus. First, we control for the effect of COVID-19 induced uncertainty. Second, for robustness purpose, we control for the effect of US EPU. To this end, we use the PWC approach allowing to measure the co-movements between oil and stock prices by controlling for the uncertainty effects. The results are presented in Figure 2 and show that, if the COVID-19 related uncertainty is defined by the log-difference of total infection cases (Figure 2a), the co-movements between WTI oil prices and S&P 500 price returns are significant in the short and medium runs, for specific moments in time. The previous WTC results are thus confirmed. Nevertheless, we observe that the yellow surfaces delimitated by the black contour (5% significance level) are smaller than the co-movement surface identified in Figure 1. Put it differently, when the effect of COVID-19 uncertainty is isolated, we observe that the co-movements manifest for a smaller period. Therefore, we conclude that COVID-19 crisis has amplified the oil-stock prices co-movements.

Fig. 2. Partial wavelet coherence between WTI and S&P 500

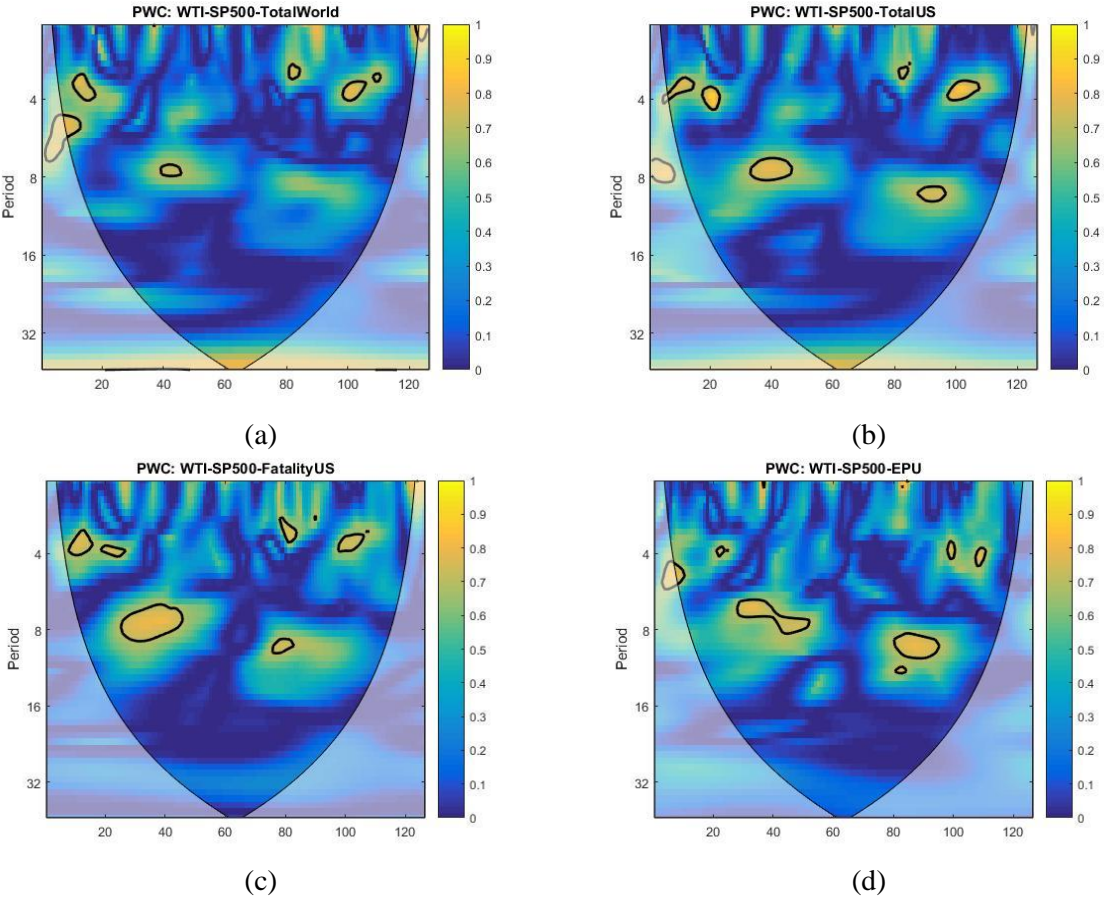

Notes: (i) On the X-axis, observation 1 corresponds to February 18, 2020, observation 60 corresponds to May 12, and observation 120 to August 6; (ii) Y-axis measures frequencies or scales (period or periodicity on the graph).

A similar result is obtained when the US infection cases are considered as a proxy for the COVID-19 uncertainty (Figure 2b). We also notice that COVID-19 fatality rate induces an uncertainty which affects the oil– stock price relationship (Figure 2c). Finally, when we isolate the effects of US EPU (Figure 2d), we notice that the co-movements manifest for a shorter period in July-August 2020 (3-5-day cycles).

# 4. Conclusions and policy implications

Several studies already tried to explain the oil – stock price nexus considering the impact of US economic policy uncertainty. Our paper contributes to this narrow strand of the literature and investigates the role of the uncertainty induced by the COVID-19 crisis. During the recent pandemic crisis, the oil price collapsed and the shocks spillover from on market to another are apparently amplified.

Our wavelets results indicate the presence of non-linear co-movement, which manifests at short to medium frequencies (2-6-day cycles) starting from February 20 to March 10, 2020, and between June and July 2020. During these periods, it is not clear if oil prices are leading stock prices, or the opposite applies. Very importantly, the developments of COVID-19 generate a significant increase in the co-movements both in the short and medium runs, whereas the US EPU amplifies the co-movements at 3-5-day cycle. These findings remain robust to different specifications of COVID-19 associated uncertainty. In terms of policy implications, the coronavirus crisis seems to be a quite cataclysmic global shock with extremely largely sanitary and economic consequences. Therefore, the economic and financial governance need to be revisited. A thorough understanding of the interactions between commodities and financial markets, and more particularly between oil and stock prices could contribute to effectively guide the policymakers' decisions.